\documentclass{article}

\usepackage{microtype}
\usepackage{graphicx}
\usepackage{subfigure}
\usepackage{booktabs} 
\usepackage{amsmath} 
\usepackage{amssymb}
\usepackage{placeins}

\usepackage[hyphens]{xurl}
\usepackage{breakurl}
\usepackage[breaklinks]{hyperref}
\usepackage{placeins}

\usepackage[accepted]{icml2019}

\begin{document}

\twocolumn[
\title{The EnvDesign Model: A Method to Solve the Environment Design Problem}
\date{\vspace{-0.2in}}
\maketitle



\icmlsetsymbol{equal}{*}

\begin{icmlauthorlist}
\icmlauthor{Akshay Sathiya, Azure Core Insights Data Science, asathiya@microsoft.com}{equal}
\icmlauthor{Rohit Pandey, Azure Core Insights Data Science, ropandey@microsoft.com}{equal}
\end{icmlauthorlist}
\vspace{0.4in}

\begin{abstract}
Today, several people and organizations rely on cloud platforms. The reliability of cloud platforms depends heavily on the performance of their internal programs (agents). To better prevent regressions in cloud platforms, the design of pre-production testing environments (that test new agents, new hardwares, and other changes) must take into account the diversity of server/node properties (hardware model, virtual machine type, etc.) across the fleet and dynamically emphasize or de-emphasize the prevalence of certain node properties based on current testing priorities. This paper formulates this task as the ``environment design" problem and presents the EnvDesign model, a method that uses graph theory and optimization algorithms to solve the environment design problem. The EnvDesign model was built on context and techniques that apply to combinatorial testing in general, so it can support combinatorial testing in other domains. An earlier version of this paper was peer-reviewed and published internally at Microsoft.
\\

\end{abstract}
\vspace{0.4in}
]

\printAffiliationsAndNotice{\icmlEqualContribution} 
\clearpage

\section{Introduction}

In recent years, cloud computing has been one of the biggest trends in technology. Many organizations have spent considerable time, money, and effort into moving their software from on-premises machines to cloud platforms.

Cloud platforms are backed by several internal programs (agents). The quality of any cloud platform depends heavily on the testing and performance of its agents. In production, these agents need to run on several environments with different configurations of server/node properties (node configurations). Hence, it is paramount that agents are tested well, to ensure that they will perform well in production and keep the cloud platform reliable for its users.

\subsection{Agent Build Deployment at Azure}

All agent builds in Azure are deployed to the production fleet through the Azure Safe Deployment Practices (SDP) pipeline \cite{AzureSDP1}. Every build is deployed in phases, where earlier phases have less diversity of node configurations as well as lower cost and impact of failures. Signals are used to monitor the health and performance of the nodes where the new build has been deployed. If the signals indicate that a regression has occurred, then action can be taken accordingly (e.g. stopping the deployment and reverting the build version) to ensure that the faulty build does not get deployed to the rest of the fleet and impact more customers.

The first two phases of the SDP pipeline are an integration environment (called ``Stage") and the Canary regions \cite{AzureSDP1}. These environments are static and noisy, since the variety of node configurations is relatively low and new builds for multiple agents are typically deployed there at the same time. The static nature of these environments makes it difficult to test builds on a diverse set of node configurations until later phases of the SDP pipeline. Additionally, multiple new builds being deployed to these environments at the same time causes a low signal-to-noise ratio, making it difficult to catch and root cause regressions.

This means that certain regressions may not get caught in Stage or Canary, and could appear in later phases of the SDP pipeline and significantly impact customers. Hence, it is crucial to test these builds in dedicated, dynamic, and diverse environments that are tailored to current testing priorities, before they are too far along the deployment pipeline.

\subsection{Pre-Production Testing Environments}

The need to test builds early, and in dedicated environments that are dynamic, diverse, and tailored to current testing priorities, boils down to the problem of determining the optimal pre-production testing environment given a constrained set of node properties and the desired levels of prevalence/target distribution of those node properties.

This paper describes the evolution of this problem, proves that it is $NP$-hard, formulates it as a combinatorial constrained optimization problem called the ``environment design" problem, and presents the EnvDesign model (designed and developed by the authors of this paper), which solves the environment design problem.

The performance of the EnvDesign model is evaluated on real Azure production data. The EnvDesign model is currently being used to design pre-production testing environments in AzQualify, a pre-production A/B testing and regression prevention system for Azure that validates agent builds as well as hardware models, virtual machine types, and other resources in Azure. 

\subsection{Generalized Combinatorial Testing}

While the EnvDesign model was designed and developed for optimizing combinatorial testing (specifically pre-production agent build testing) in Azure, the underlying techniques can be applied to combinatorial testing in general. Hence, the EnvDesign model can be used to optimize combinatorial testing in the tech industry and beyond the tech industry.

\section{The Environment Design Problem}

\subsection{Evolution of the Problem}

The environment design problem began as a relatively easy problem but evolved into harder and harder problems as requirements inevitably piled on, with the addition of more inputs and constraints \cite{ExpDesign}.

\begin{figure}[ht]
\begin{center}
\centerline{\includegraphics[width=3.5cm]{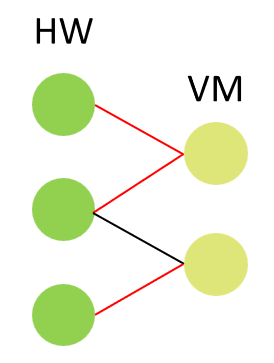}}
\caption{Environment design problem, minimum edge cover}
\label{MinEdgeCover}
\end{center}
\vskip -0.3in
\end{figure}

The simplest version of the environment design problem begins with two ``dimensions" (node properties, in the context of Azure), hardware model (HW) and virtual machine (VM) type. The ``dimension values" (specific HW models and VM types) and their compatibility relationships can be represented as a bipartite graph $G = (V = V_1 \cup V_2, E)$, which is shown in Figure \ref{MinEdgeCover}. $V_1$ is the set of vertices corresponding to HW models and $V_2$ is the set of vertices corresponding to VM types. Vertices in the same layer of the graph (dimension values in the same dimension) do not have edges/compatibility relationships with each other.

In this case, each node configuration consists of two dimension values (a HW model and a VM type) and corresponds to an edge in the graph. To ensure that the chosen node configurations are diverse, they must satisfy a coverage constraint, by collectively covering all dimension values (i.e., covering all vertices in the graph).

Finding the smallest set of node configurations needed to cover all HW models and VM types is the same as the minimum edge cover problem for bipartite graphs, which can be solved in polynomial time \cite{Schrijver}. In Figure \ref{MinEdgeCover}, the red edges correspond to the minimum edge cover.

However, testing will be more effective if more dimensions are taken into account. In practice, another important dimension to account for in testing is the guest operating system (OS) image. 

\begin{figure}[ht]
\begin{center}
\centerline{\includegraphics[width=5cm]{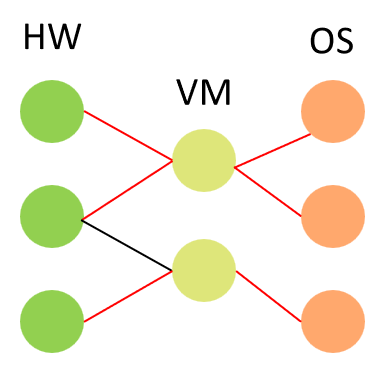}}
\caption{Environment design problem, minimum path cover}
\label{MinPathCover}
\end{center}
\vskip -0.3in
\end{figure}

For simplicity, let's assume that compatibility relationships exist only between successive dimensions/layers of the graph (HW model to VM type and VM type to OS image). This preserves the bipartite nature of the graph. The updated graph $G = (V = V_1 \cup V_2 \cup V_3, E)$ is shown in Figure \ref{MinPathCover}. $V_3$ is the set of vertices corresponding to OSes. 

In this case, each node configuration consists of three dimension values (one HW model, one VM type, and one OS image) and correspond to paths of length 3 in the graph (with one vertex from each layer of the graph).

The problem of finding the smallest set of node configurations needed to cover all HW models, VM types, and OS images is essentially the minimum path cover problem for bipartite graphs, which can also be solved in polynomial time \cite{ExpDesign}. In Figure \ref{MinPathCover}, the red edges and the vertices connected by them correspond to the minimum path cover. 

\begin{figure}[ht]
\begin{center}
\centerline{\includegraphics[width=5cm]{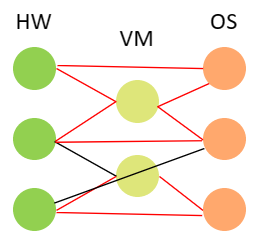}}
\caption{Environment design problem, minimum clique cover}
\label{MinCliqueCover}
\end{center}
\vskip -0.3in
\end{figure}

However, when compatibility relationships exist between all dimensions, not just between successive ones, the graph will no longer be bipartite. The updated sample graph $G$ is shown in Figure \ref{MinCliqueCover}.

In this case, each node configuration corresponds to a clique of size 3 in the graph. The problem of finding the minimal set of node configurations needed to cover each HW model, VM type, and OS image is a more constrained version of the minimum clique cover problem (where each clique must include a vertex from each layer of the graph).

The clique cover problem is an $NP$-complete decision problem that asks if there exists a collection of cliques, the size of which is no greater than a given bound, that covers all the vertices in a given graph \cite{Karp}. By definition, all $NP$-complete problems are also $NP$-hard \cite{CLR4}, so the clique cover problem is also $NP$-hard. Since the clique cover problem is $NP$-hard, it likely cannot be solved in polynomial time \cite{Optimizn}.

The minimum clique cover problem is an optimization problem, where the optimal solution is the clique cover consisting of the smallest number of cliques. Any optimization problem is at least as hard as its corresponding decision problem \cite{CLR4}, which in this case is the $NP$-hard clique cover problem. It follows that the minimum clique cover problem is also $NP$-hard and the optimal solution likely cannot be obtained in polynomial time. 

Hence, the environment design problem has now evolved into a more constrained version of the $NP$-hard minimum clique cover problem.

In Figure \ref{MinCliqueCover}, the red edges and the vertices connected by them correspond to the graph's minimum clique cover.

In practice, we often have a specified number of nodes for testing, $n$. It follows that the size of the clique cover must be less than or equal to $n$. Now that the size of the clique cover is bounded, the environment design problem has become a more constrained version of the $NP$-hard clique cover problem (the decision problem with a bound, as opposed to the minimum clique cover problem, an optimization problem without a bound).

Although all $n$ nodes may not be needed to satisfy the coverage constraint, it is best to use all $n$ nodes and perform as much testing as possible. The configurations for these $n$ nodes must be determined in such a way that the coverage constraint is still satisfied and, to the greatest extent possible, the desired/target distribution of dimension values is matched as closely as possible in the collection of testing configurations (schedule). Let this closeness between true and target distributions be quantified by an objective function $O(S)$, where $S$ is a schedule. When $O(S) = 0$, the distribution matches exactly with the target distributions.

For some testing scenarios, only some of the dimension values in a dimension need to be covered and other dimension values need to be excluded. These constraints are represented by a scope $C = (I = I_1 \cup I_2 \cup I_3, X = X_1 \cup X_2 \cup X_3)$, where $I$ contains the include scope (dimension values in each dimension that need to be covered) and $X$ contains the exclude scope (dimension values in each dimension that need to be excluded). We assume that the two sets don't intersect for any dimension (if they did, the problem would be immediately infeasible since a dimension value cannot be included and excluded at the same time).

Since the environment design problem is a more constrained version of the $NP$-hard clique cover problem, is an $NP$-hard problem itself (proof provided in section \ref{EDP_NPHard_Proof} of this paper), and seeks to minimize the value of an objective function, it is an $NP$-hard constrained optimization problem.

\subsection{Formulation of the Environment Design Problem}

While many constrained optimization problems can be formulated with decision variables, the environment design problem is not formulated as such.

Solutions to the environment design problem are collections of $n$ cliques, where each clique is of size $d$ and contains one vertex from each of the $d$ dimensions ($V_1, \dots, V_d$).

Suppose that solutions to the environment design problem are expressed through decision variables. There are at most (for dense graphs) $\prod_{i=1}^d |V_i|$ cliques that need to be considered. So, there would be at most $\prod_{i=1}^d |V_i|$ decision variables, where each decision variable corresponds to a clique and its respective value is the number of times the corresponding clique is used in the solution. To maintain a mapping between these decision variables and their corresponding cliques, all of the cliques need to be enumerated at once and stored in memory. Optimization algorithms developed under this expression of the problem would risk running into out-of-memory errors and would waste lots of time at the start of their execution enumerating cliques, rather than exploring different solutions.

Now, suppose that solutions to the environment design problem are expressed as length-$n$ lists of cliques, where each element in the list is a size-$d$ clique. Simulated annealing algorithms developed under this expression of the problem can modify a given list of cliques by iteratively replacing a single clique in the list with another clique. For branch and bound algorithms developed under this expression of the problem, each branching step could correspond to adding one of $b$ (a specified branch factor) cliques to a partially-assembled list of cliques, the length of which is less than $n$.

In these simulated annealing and branch and bound algorithms, cliques can be produced as they are needed, rather than all at once. Additionally, for the branch and bound algorithms, the branch factor $b$ can be used to control the number of cliques produced at each branching step. These algorithms are less at risk of running into out-of-memory errors and can explore different solutions much earlier in their execution. These algorithms are discussed in greater detail in section \ref{Optimization} of this paper. 

Hence, the environment design problem is formulated in such a way that solutions are lists of cliques, rather than configurations of decision variables. The formulation of the environment design problem as a constrained optimization problem is shown below.

Inputs:
\begin{enumerate}
\item An undirected graph $G = (V = V_1 \cup \dots \cup V_d, E)$ where the vertices in $V$ are split into $d \geq 2$ groups, each group representing a testing dimension (e.g. HW model, VM type) and each vertex representing a dimension value. The edges in $E$ represent compatibility relationships between the vertices in $V$. Dimension values in the same dimension are considered incompatible $\forall k \in \{1, \dots, d\}, \forall v_i, v_j \in V_k, (v_i, v_j) \not\in E$
\item The number of testing configurations $n \in \mathbb{Z}^+$
\item Objective function $O(S)$, that when given a collection of testing configurations (``schedule") $S$, returns a non-negative number that quantifies how close the target distribution of dimension values is to the distribution of dimension values in $S$ (0 indicates a perfect match). 
\item Scope $C = (I = I_1 \cup \dots \cup I_d, X = X_1 \cup \dots \cup X_d)$, $\forall i, I_i \cap X_i = \emptyset$, which consists of collections of dimension values (for each dimension) that must be included in ($I$) and excluded from ($X$) the output.
\end{enumerate}

Output: 
\begin{itemize}
\item Schedule $S = [s_1, s_2, \dots, s_n]$
\end{itemize}

Constraints:
\begin{enumerate}
\item Schedule contains $n$ testing configurations, $|S| = n$
\item Each testing configuration has $d$ dimension values $\forall s \in S, |s| = d$, exactly one dimension value from each dimension $\forall s \in S, \forall i \in \{1, \dots, d\}, |s \cap V_i| = 1$
\item Dimension values in each testing configuration must all be compatible with each other $\forall s \in S, \forall s_i, s_j \in s,$ where $i \neq j$, $(s_i, s_j) \in E$
\item Schedule does not contain any dimension values in the exclude scope, $\forall s \in S, \forall i \in \{1, \dots, d\}, s \cap X_i = \emptyset$
\item Schedule covers every dimension value in the include scope, for dimensions where it is provided $\forall i \in \{1, \dots, d\}$ where $I_i \neq \emptyset, \forall v \in V_i \cap I_i, \exists s \in S, v \in s$
\item Schedule does not cover dimension values outside the include scope, for dimensions where it is provided, $\forall i \in \{1, \dots, d\}$ where $I_i \neq \emptyset, \forall v \in V_i - I_i, \nexists s \in S, v \in s$
\end{enumerate}

Optimize:
\begin{itemize}
\item Minimize the value of the objective function, $O(S)$
\item Optimal solution is $\text{min}_S \{O(S)\}$
\end{itemize}

An example instance of the environment design problem, and the optimal solution, is shown in section \ref{ExampleInstance} of this paper. 

\section{EnvDesign Model}

The EnvDesign model produces schedules through three main steps: graph algorithms, optimization algorithms, and postprocessing. Each of these steps are described in greater detail below.

\subsection{Graph Algorithms}

Graph algorithms are used to satisfy the scope constraints and the coverage constraint.

First, the graph is ``scoped". That is, vertices and edges are removed from $G$ based on the scope $C = (I = I_1 \cup I_2 \cup I_3, X = X_1 \cup X_2 \cup X_3)$. For each dimension $i \in {1, \dots, d}$, the values in the corresponding exclude scope $X_i$ are removed from the graph, $V_i = V_i - X_i$. Additionally, if an include scope is provided, $I_i \neq \emptyset$, all vertices outside the include scope are removed from the graph, $V_i = V_i \cap I_i$.

In some cases, the graph can still be quite large, even after scoping. To improve the performance of the following algorithms, the graph size can be reduced.
If vertices in $V - I$ do not have edges to every other layer of the graph, they cannot be used to form cliques can be removed. Additionally, if $I \neq \emptyset$, all vertices that are not in $I$ and do not have an edge to a vertex in $I$ can be removed, since they cannot be used to produce cliques under the given scope.

So, a clique cover of the graph is obtained using a modified version of the Tomita algorithm presented in \cite{Tomita} (the ``build-cliques" algorithm). Starting from a given vertex, this algorithm performs a kind of depth-first-search (where each successive vertex neighbors all the previous vertices) to add more vertices to the clique until it reaches the desired size. This algorithm is run from every uncovered vertex (starting with the vertices in $V \cap I$) and the resulting cliques are collected to obtain the clique cover $Q$. In the depth-first-search, vertices that have not yet been covered by a clique are considered before vertices that have already been covered by a clique(s).

Also, each layer of the graph can be restricted to a certain size (maximum dimension size). The vertices with the highest levels of prevalence in the target distribution can be kept, and others can be removed. If $I \neq \emptyset$, then this can be done after all vertices in $V \cap I$ have been covered by cliques. Otherwise, this can be done before obtaining the clique cover.

Additionally, if certain vertices cannot be covered by cliques, they can be removed from the graph if they are not in $I$.

The coverage constraint now only pertains to the remaining vertices in the graph. The clique cover $Q$ corresponds to a coverage schedule, which satisfies the coverage constraint and the scope constraints.

\subsection{Optimization} \label{Optimization}

If $|Q| < n$, then $Q$ (and/or a portion of $Q$) are duplicated until an expanded coverage schedule $S_0$ is formed, where $|S_0| = n$. The expanded coverage schedule $S_0$ is the initial schedule that will be optimized under the objective function $O(S)$.

Using the optimizn library (developed by the authors of this paper), simulated annealing and branch and bound algorithms were developed for optimizing the initial expanded coverage schedule $S_0$. The optimizn library and its offerings (simulated annealing, branch and bound, and continuous training) are discussed in greater detail in other literature written by the authors of this paper \cite{Optimizn}. The Microsoft open-source repository for the optimizn library is located here: \url{https://github.com/microsoft/optimizn}.

The simulated annealing algorithm iteratively modifies individual cliques in $S_0$ to produce a new, more optimal schedule $S$. This customizable components of the simulated annealing algorithm in optimizn are defined as follows. 
\begin{itemize}
\itemsep 0em
\item \texttt{get\_initial\_solution}: Returns $S_0$.
\item \texttt{reset\_candidate}: Returns a random schedule $S_R$ that satisfies the coverage constraint, which contains $Q$ and $n - |Q|$ randomly selected cliques from $Q$.
\item \texttt{next\_candidate}: Replaces a randomly-chosen clique $s \in S$ with a clique $t \neq s$.

If the coverage constraint is not satisfied by the remaining cliques, $t$ is created by the build-cliques algorithm starting from the vertices in $s$ that are not covered by the other cliques.

If the coverage constraint is satisfied by the remaining cliques, then $t$ is created in one of the following ways.
\begin{itemize}
    \item Created by the build-cliques algorithm starting from a randomly selected vertex, not necessarily in $s$.
    \item Created by the build-cliques algorithm starting from all but one of the vertices in $s$.
    \item Created by the build-cliques algorithm starting from a single vertex $v \in s$.
\end{itemize}

If a replacement clique $t \neq s$ that preserves the coverage constraint cannot be found, then this process of randomly choosing and replacing a clique in $S$ is repeated some number of times (the retries limit). If a replacement clique is not found by the time the retries limit is reached, then the \texttt{reset\_candidate} function is used to produce a new schedule. Alternatively, the chosen clique $s$ could be chosen from $S - Q$ (which only contains repeats of $Q$ and/or repeated portions of $Q$), which guarantees preservation of the coverage constraint.
\item \texttt{cost}: Returns $O(S)$. In the EnvDesign model, $O(S)$ can be one of the following.
\begin{itemize}
    \item Dimension-based: computes the weighted sum of mean-squared-errors between the target/true distributions of dimension values in each dimension
    \item Relationship-based: computes weighted the sum of mean-squared-errors between the target/true distributions for relationships (pairs of dimension values) in each pair of dimensions
    \item Combination-based: computes the mean-squared-error between the target/true distributions of dimension value configurations (spanning all dimensions)
\end{itemize}

If dimension values were removed from the graph earlier, then the target distributions must be adjusted so they only pertain to remaining dimension values (for dimension-based objective functions), relationships between remaining dimension values (for relationship-based objective functions) or combinations of the remaining dimension values (for combination-based objective functions).

The target distributions are adjusted by first discarding the target distribution values that do not pertain to a remaining dimension value, relationship between remaining dimension values, or combination of remaining dimension values. Then, for combination-based objective functions, the remaining target distribution values are normalized. For dimension-based objective functions, the remaining target distribution values are normalized by dimension. For relationship-based objective functions, the remaining target distribution values are normalized by pair of dimensions.

\item \texttt{get\_temperature}: Returns the output of the following function, where $x$ is the number of iterations since the last random restart.
$$f(x) = \frac{4000}{1 + e^{x / 3000}}$$
\end{itemize}

The first branch and bound algorithm builds a schedule $S$, clique by clique, starting from an empty collection of cliques. This customizable components of the branch and bound algorithm in optimizn are defined as follows. 
\begin{itemize}
\itemsep 0em
\item \texttt{get\_initial\_solution}: Same as simulated annealing algorithm.
\item \texttt{get\_root}: Returns an empty schedule, $[]$
\item \texttt{lbound}: For a given schedule $S$, if $O(S)$ is dimension-based, then for each dimension, the values of the true and target distributions are multiplied by $n$ to get the true and target numbers of nodes for each dimension value in that dimension. Then, the true number of nodes for the dimension value with the largest difference between target and true numbers of nodes is incremented by 1. This is done $n - |S|$ times. Then, for each dimension, the target and true numbers of nodes of each dimension value are divided by $n$ (normalized) to get the target distribution and updated true distribution, from which $O(S)$ is calculated to get the lowest possible cost of a length-$n$ schedule produced from $S$ (the lower bound).

Similarly, if $O(S)$ is relationship-based, then this is done with respect to the relationships for each pair of dimensions. If $O(S)$ is combination-based, then this is done with respect to node configurations.

Essentially, the lower bound is calculated by iteratively and greedily determining the node configurations for the $n - |S|$ additional nodes (which need not be valid node configurations), such that the value of $O(S)$ is decreased by the largest possible amount or increased by the smallest possible amount in each iteration.
\item \texttt{cost}: Same as simulated annealing algorithm.
\item \texttt{is\_feasible}: Checks if a given schedule $S$ satisfies the coverage constraint and if $|S| = n$.
\item \texttt{complete\_solution}: Iteratively adds cliques from $Q$ to a given schedule $S$, until $|S| = n$. In each iteration, if all vertices are covered, the added clique is chosen randomly from all cliques in $Q$. Otherwise, the added clique is chosen greedily (covers the most uncovered vertices) from all cliques in $Q$.
\item \texttt{branch}: Given a schedule $S$ (where $|S| < n$) and a branch factor $b$, returns $b$ schedules, each of which is $S_P$ with an additional clique that covers at least one uncovered vertex. If all vertices are covered by $S_P$, then additional cliques can be any cliques in the graph.
\end{itemize}

The second branch and bound algorithm modifies cliques in $S_0$ to produce a new, more optimal schedule $S$. 
\begin{itemize}
\itemsep 0em
\item \texttt{get\_initial\_solution}: Same as previous algorithms.
\item \texttt{get\_root}: Same as previous branch and bound algorithm.
\item \texttt{lbound}: Same as previous branch and bound algorithm.
\item \texttt{cost}: Same as previous algorithms.
\item \texttt{is\_feasible}: Same as previous branch and bound algorithm.
\item \texttt{complete\_solution}: Adds the last $n - |S|$ cliques from $S_0$ to a given schedule $S$.
\item \texttt{branch}: Given a schedule $S$ (where $|S| < n$) and a branch factor $b$, returns $b$ new schedules, each of which is $S$ with an additional clique that preserves the coverage constraint when followed by the last $n - |S| - 1$ cliques from $S_0$ .
\end{itemize}

\subsection{Postprocessing}

The resulting schedule $S$ contains $n$ node configurations, each contianing a VM type. If the schedule is run in its current form, then $n$ VMs will be used for testing.

However, in practice, nodes can be ``packed" with multiple VMs. It would be best to pack each node with as many VMs as possible to perform the most amount of testing possible. 

Hence, each testing configuration $s \in S$ can be duplicated $w_s - 1$ times, where $w_s$ is the maximum number of VMs (of the VM type specified in $s$) that can be run on the node (with the HW model specified in $s$). These $w_s$ testing configurations would be run on the same node.

\section{Experiments}

Two experiments are run to evaluate the EnvDesign model, both of which test the EnvDesign model's performance on two instances of the environment design problem using real Azure data.

\subsection{System Specifications}

Both experiments are run on their own Azure Databricks clusters, each with the following specifications.
\begin{itemize}
\itemsep 0em 
\item VMSKU: Standard\_DS3\_v2
\item Nodes: 1
\item Workers/Drivers: 1
\item Memory: 14 GB
\item Cores: 4
\item Databricks Runtime Version: 13.3 LTS
\end{itemize}

\subsection{Optimization Algorithms}

In both experiments, the following optimization algorithms (6 simulated annealing, 12 branch and bound) are tested on different instances of the environment design problem.

The retries limit for all simulated annealing algorithms is 8, and the reset probability is 1e-7. The branch factor for branch and bound algorithms with a depth-first or depth-first-best-first selection strategy is 500. The branch factor for branch and bound algorithms with a best-first-depth-first selection strategy is 50, since branch and bound under this selection strategy is less space-efficient than under the depth-first or depth-first-best-first strategies \cite{Optimizn}.

\begin{itemize}
\item 1.1: simulated annealing to optimize the expanded coverage schedule (without preserving clique cover, replacing each chosen clique with a clique built from a random vertex)
\item 1.2: simulated annealing to optimize the expanded coverage schedule (with preservation of clique cover, replacing each chosen clique with a clique built from a random vertex)
\item 1.3: simulated annealing to optimize the expanded coverage schedule (without preserving clique cover, replacing each chosen clique with a clique built from all but one of the vertices in the chosen clique)
\item 1.4: simulated annealing to optimize the expanded coverage schedule (with preservation of clique cover, replacing each chosen clique with a clique built from all but one of the vertices in the chosen clique)
\item 1.5: simulated annealing to optimize the expanded coverage schedule (without preserving clique cover, replacing each chosen clique with a clique built from a single vertex in the chosen clique)
\item 1.6: simulated annealing to optimize the expanded coverage schedule (with preservation of clique cover, replacing each chosen clique with a clique built from a single vertex in the chosen clique)
\item 2.1: traditional branch and bound to generate full schedule from scratch, with depth-first selection strategy
\item 2.2: look-ahead branch and bound to generate full schedule from scratch, with depth-first selection strategy
\item 2.3: traditional branch and bound to generate full schedule from scratch, with depth-first-best-first selection strategy
\item 2.4: look-ahead branch and bound to generate full schedule from scratch, with depth-first-best-first selection strategy
\item 2.5: traditional branch and bound to generate full schedule from scratch, with best-first-depth-first selection strategy
\item 2.6: look-ahead branch and bound to generate full schedule from scratch, with best-first-depth-first selection strategy
\item 3.1: traditional branch and bound to optimize expanded coverage schedule, with depth-first selection strategy
\item 3.2: look-ahead branch and bound to optimize the expanded coverage schedule, with depth-first selection strategy
\item 3.3: traditional branch and bound to optimize expanded coverage schedule, with depth-first-best-first selection strategy
\item 3.4: look-ahead branch and bound to optimize expanded coverage schedule, with depth-first-best-first selection strategy
\item 3.5: traditional branch and bound to optimize expanded coverage schedule, with best-first-depth-first selection strategy
\item 3.6: look-ahead branch and bound to optimize expanded coverage schedule, with best-first-depth-first selection strategy
\end{itemize} 

\subsection{Experiment 1}

In the first experiment, the testing dimensions are HW model, basic input/output system (BIOS) version, and VM type. The desired testing schedule must consist of 150 nodes. The dimension values and compatibility relationships are obtained from a sample of 1000 real node configurations observed in a 30-day snapshot of Azure (February 25th to March 25th, 2024).

The EnvDesign model is run three times in succession, with continuous training (each run picks up from where the previous run left off). In each run, the optimization algorithms are given 30 minutes of compute time.

The runs are performed for each type of objective function (dimension-based, relationship-based, and combination-based), where the target and true distributions for each dimension value, relationship, and dimension value configuration is its count (the number of times it appeared in the sample).

The optimality of the schedules produced by each algorithm across its runs will be compared.

The results of Experiment 1, under the dimension-based, relationship-based, and combination-based objective functions, are shown in Tables \ref{Exp1ResultsDim}, \ref{Exp1ResultsRel}, and \ref{Exp1ResultsComb} respectively.

\begin{table}[t]
\vskip 0in
\begin{center}
\begin{small}
\begin{sc}
\begin{tabular}{lcccc}
\toprule
Opt. Alg. & Init. & Run 1 & Run 2 & Run 3 \\
\midrule
1.1 & 2.197e-3 & 2.044e-3 & 2.044e-3 & 2.002e-3 \\
1.2 & 2.197e-3 & 1.799e-3 & 1.799e-3 & 1.799e-3 \\
1.3 & 2.197e-3 & 1.817e-3 & 1.817e-3 & 1.817e-3 \\
1.4 & 2.197e-3 & 2.070e-3 & 2.070e-3 & 2.070e-3 \\
1.5 & 2.197e-3 & 2.085e-3 & 2.085e-3 & 1.944e-3 \\
1.6 & 2.197e-3 & 2.197e-3 & 2.197e-3 & 2.197e-3 \\
2.1 & 2.197e-3 & 2.109e-3 & 2.084e-3 & 2.072e-3 \\
2.2 & 2.197e-3 & 2.105e-3 & 2.084e-3 & 2.072e-3 \\
2.3 & 2.197e-3 & 2.197e-3 & 2.197e-3 & 2.197e-3 \\
2.4 & 2.197e-3 & 2.017e-3 & 2.017e-3 & 2.017e-3 \\
2.5 & 2.197e-3 & 2.197e-3 & 2.197e-3 & 2.197e-3 \\
2.6 & 2.197e-3 & 2.150e-3 & 2.111e-3 & 2.111e-3 \\
3.1 & 2.197e-3 & 2.002e-3 & 1.981e-3 & 1.970e-3 \\
3.2 & 2.197e-3 & 2.002e-3 & 1.981e-3 & 1.970e-3 \\
3.3 & 2.197e-3 & \textbf{1.534e-3} & \textbf{1.534e-3} & \textbf{1.534e-3} \\
3.4 & 2.197e-3 & \textbf{1.534e-3} & \textbf{1.534e-3} & \textbf{1.534e-3} \\
3.5 & 2.197e-3 & 2.197e-3 & 2.197e-3 & 2.197e-3 \\
3.6 & 2.197e-3 & 2.141e-3 & 2.120e-3 & 2.118e-3 \\
\bottomrule
\end{tabular}
\end{sc}
\end{small}
\end{center}
\caption{Optimality of testing schedules under a dimension-based objective function (Experiment 1)}
\vskip 0in
\label{Exp1ResultsDim}
\end{table}

\begin{table}[t]
\vskip 0in
\begin{center}
\begin{small}
\begin{sc}
\begin{tabular}{lcccc}
\toprule
Opt. Alg. & Init. & Run 1 & Run 2 & Run 3 \\
\midrule
1.1 & 2.823e-4 & 2.546e-4 & 2.520e-4 & 2.520e-4 \\
1.2 & 2.823e-4 & 2.703e-4 & 2.690e-4 & 2.686e-4 \\
1.3 & 2.823e-4 & \textbf{2.510e-4} & \textbf{2.509e-4} & \textbf{2.509e-4} \\
1.4 & 2.823e-4 & 2.650e-4 & 2.650e-4 & 2.650e-4 \\
1.5 & 2.823e-4 & 2.568e-4 & 2.568e-4 & 2.568e-4 \\
1.6 & 2.823e-4 & 2.769e-4 & 2.769e-4 & 2.769e-4 \\
2.1 & 2.823e-4 & 2.805e-4 & 2.789e-4 & 2.788e-4 \\
2.2 & 2.823e-4 & 2.720e-4 & 2.716e-4 & 2.716e-4 \\
2.3 & 2.823e-4 & 2.823e-4 & 2.823e-4 & 2.823e-4 \\
2.4 & 2.823e-4 & 2.768e-4 & 2.758e-4 & 2.614e-4 \\
2.5 & 2.823e-4 & 2.823e-4 & 2.823e-4 & 2.823e-4 \\
2.6 & 2.823e-4 & 2.823e-4 & 2.823e-4 & 2.823e-4 \\
3.1 & 2.823e-4 & 2.698e-4 & 2.678e-4 & 2.644e-4 \\
3.2 & 2.823e-4 & 2.698e-4 & 2.678e-4 & 2.644e-4 \\
3.3 & 2.823e-4 & 2.823e-4 & 2.823e-4 & 2.823e-4 \\
3.4 & 2.823e-4 & 2.768e-4 & 2.683e-4 & 2.598e-4 \\
3.5 & 2.823e-4 & 2.823e-4 & 2.823e-4 & 2.823e-4 \\
3.6 & 2.823e-4 & 2.768e-4 & 2.768e-4 & 2.768e-4 \\
\bottomrule
\end{tabular}
\end{sc}
\end{small}
\end{center}
\caption{Optimality of testing schedules under a relationship-based objective function (Experiment 1)}
\vskip 0in
\label{Exp1ResultsRel}
\end{table}

\begin{table}[t]
\vskip 0in
\begin{center}
\begin{small}
\begin{sc}
\begin{tabular}{lcccc}
\toprule
Opt. Alg. & Init & Run 1 & Run 2 & Run 3 \\
\midrule
1.1 & 5.255e-5 & 4.917e-5 & 4.917e-5 & 4.917e-5 \\
1.2 & 5.255e-5 & 5.001e-5 & 5.001e-5 & 5.001e-5 \\
1.3 & 5.255e-5 & 5.013e-5 & 5.013e-5 & 5.013e-5 \\
1.4 & 5.255e-5 & 5.042e-5 & 5.042e-5 & 5.042e-5 \\
1.5 & 5.255e-5 & 5.066e-5 & 5.066e-5 & 5.066e-5 \\
1.6 & 5.255e-5 & 5.167e-5 & 5.167e-5 & 5.167e-5 \\
2.1 & 5.255e-5 & 4.742e-5 & 4.518e-5 & 4.307e-5 \\
2.2 & 5.255e-5 & 4.742e-5 & 4.491e-5 & 4.292e-5 \\
2.3 & 5.255e-5 & 5.255e-5 & 5.255e-5 & 5.255e-5 \\
2.4 & 5.255e-5 & 4.528e-5 & 4.390e-5 & 4.390e-5 \\
2.5 & 5.255e-5 & 5.255e-5 & 5.255e-5 & 5.255e-5 \\
2.6 & 5.255e-5 & 4.998e-5 & 4.946e-5 & 4.882e-5 \\
3.1 & 5.255e-5 & \textbf{4.492e-5} & 4.271e-5 & 4.155e-5 \\
3.2 & 5.255e-5 & \textbf{4.492e-5} & 4.271e-5 & 4.155e-5 \\
3.3 & 5.255e-5 & 5.255e-5 & \textbf{3.992e-5} & \textbf{3.992e-5} \\
3.4 & 5.255e-5 & 4.522e-5 & \textbf{3.992e-5} & \textbf{3.992e-5} \\
3.5 & 5.255e-5 & 5.255e-5 & 5.255e-5 & 5.255e-5 \\
3.6 & 5.255e-5 & 5.017e-5 & 4.970e-5 & 4.938e-5 \\
\bottomrule
\end{tabular}
\end{sc}
\end{small}
\end{center}
\caption{Optimality of testing schedules under a combination-based objective function (Experiment 1)}
\vskip 0in
\label{Exp1ResultsComb}
\end{table}

\subsection{Experiment 2}

In the second experiment, the instance of the environment design problem corresponds the Overlake scenario, a real testing scenario in Azure that pertains to new hardware models. 

The testing dimensions are HW model, hypervisor generation, VM type, OS image, and workload. The dimension values and relationships (for all dimensions except workload) are obtained from a 30-day snapshot of Azure (March 13th to April 11th, 2024) and from cache tables that contain compatibility relationships observed in Azure over a longer period of time. Workloads and their compatibility relationships are obtained from AzQualify database tables. The maximum dimension size of the graph is 100. The desired number of nodes is 300 and the objective function is dimension-based. The Overlake scenario scope contains hardware models and workloads that must be covered in the schedules.

The EnvDesign model is run three times in succession (with continuous training, one hour of compute time for each run) to produce and improve schedules for the Overlake scenario. The optimality of the schedules produced by each algorithm across its runs will be compared.

The results of Experiment 2 are shown in Table \ref{Exp2Results}.

\begin{table}[t]
\vskip 0in
\begin{center}
\begin{small}
\begin{sc}
\begin{tabular}{lcccc}
\toprule
Opt. Alg. & Init. & Run 1 & Run 2 & Run 3 \\
\midrule
1.1 & 3.682e-2 & 2.162e-3 & 2.156e-3 & 2.156e-3 \\
1.2 & 3.682e-2 & \textbf{1.647e-3} & \textbf{1.647e-3} & \textbf{1.647e-3} \\
1.3 & 3.682e-2 & 1.347e-2 & 1.347e-2 & 1.347e-2 \\
1.4 & 3.682e-2 & 1.258e-2 & 1.258e-2 & 1.258e-2 \\
1.5 & 3.682e-2 & 2.239e-2 & 2.239e-2 & 2.239e-2 \\
1.6 & 3.682e-2 & 2.265e-2 & 2.265e-2 & 2.265e-2 \\
2.1 & 3.682e-2 & 3.682e-2 & 3.682e-2 & 3.682e-2 \\
2.2 & 3.682e-2 & 3.183e-2 & 3.183e-2 & 3.183e-2 \\
2.3 & 3.682e-2 & 3.604e-3 & 3.604e-3 & 3.604e-3 \\
2.4 & 3.682e-2 & 3.569e-3 & 3.568e-3 & 3.568e-3 \\
2.5 & 3.682e-2 & 3.682e-2 & 3.682e-2 & 3.682e-2 \\
2.6 & 3.682e-2 & 2.814e-2 & 2.767e-2 & 2.767e-2 \\
3.1 & 3.682e-2 & 3.505e-2 & 3.505e-2 & 3.505e-2 \\
3.2 & 3.682e-2 & 3.505e-2 & 3.505e-2 & 3.505e-2 \\
3.3 & 3.682e-2 & 3.694e-3 & 3.694e-3 & 3.694e-3 \\
3.4 & 3.682e-2 & 3.694e-3 & 3.694e-3 & 3.694e-3 \\
3.5 & 3.682e-2 & 3.682e-2 & 3.682e-2 & 3.682e-2 \\
3.6 & 3.682e-2 & 3.636e-2 & 3.601e-2 & 3.601e-2 \\
\bottomrule
\end{tabular}
\end{sc}
\end{small}
\end{center}
\caption{Optimality of testing schedules for the Overlake scenario (Experiment 2)}
\vskip 0in
\label{Exp2Results}
\end{table}

\subsection{Analysis}

In Experiment 1, the EnvDesign model produced the most optimal schedules with optimization algorithms 3.3 and 3.4 (traditional and look-ahead (respectively) branch and bound to optimize the expanded coverage schedule, with a depth-first-best-first selection strategy) under the dimension-based and combination-based objective functions. Under relationship-based objective functions, the EnvDesign model produced the most optimal schedules with optimization algorithm 1.3 (simulated annealing to optimize the expanded coverage schedule, without preservation of clique cover and replacing each chosen clique with a clique built from all but one of the vertices in the chosen clique).

In Experiment 2, the EnvDesign model produced the most optimal schedules with optimization algorithm 1.2 (simulated annealing to optimize the expanded coverage schedule, with preservation of clique cover and replacing each chosen clique with a clique built from a random vertex).

In both experiments, the EnvDesign model produced schedules that had fairly low objective function values and were noticeably more optimal than the initial expanded coverage schedule, which is a good indication that the EnvDesign model can be used to produce near-optimal testing schedules for virtually any testing scenario in practice.

\section{Conclusion}

This paper discusses the evolution of the environment design problem, proves that it is $NP$-hard, and formulates it as a constrained optimization problem, where the goal is to dynamically provide a diverse testing schedule tailored to current testing priorities. 

This paper also presents the EnvDesign model, a model for solving the environment design problem that uses graph theory and optimization algorithms to produce and optimize schedules. The efficacy of the EnvDesign model was demonstrated on real Azure data and for a real testing scenario in Azure. The EnvDesign model is currently being used in AzQualify to produce pre-production testing schedules for agent builds in Azure. While the EnvDesign model was developed for agent build testing in Azure, its concepts apply to any domain and can be used to power combinatorial testing beyond Azure.

To improve the EnvDesign model, we have added support for availability constraints (i.e. maximum number of nodes that can have a particular dimension value), so we can take into account hardware model availability constraints in AzQualify. Since we intend to discuss this work in future literature, availability constraints and the updates made to the EnvDesign model to take them into account were not discussed in this paper and were not used in this paper's experiments. We intend to add support for heterogeneous nodes (nodes packed with different types of VMs, which can have different OS images and/or workloads) and support any other functionality based on needs that arise in practice. 

\section{Acknowledgements}
We would like to acknowledge Microsoft and Microsoft Azure for giving us the opportunity to build and showcase the EnvDesign model as well as apply it to our work. We would also like to acknowledge the reviewers at Microsoft for their feedback and help during the paper's review/revision. We would like to thank Chavi Gupta and Mohammad Ali Bashiri for suggesting a depth-first search algorithm for adding vertices to a clique to produce a larger clique of the desired size. Algorithm development and literature review done along the lines of their suggestion led to a modified version of the Tomita algorithm, the build-cliques algorithm, which is used in the EnvDesign model. We also want to thank Lokesh Dogga, Abhishek Kumar Singh, and Venkata Sumanth Reddy Kota for helping design the input and output contracts for the EnvDesign model and helping integrate the EnvDesign model with AzQualify. We would like to thank Anubhav Natani and the AzQualify engineering team as well for helping integrate the EnvDesign model with AzQualify. We would like to thank Gopal Jaiswal for providing a database query to get hardware model availability data and for measuring the impact in AzQualify of taking hardware model availability constraints into account.

\newpage 
\bibliographystyle{ieeetr}
\bibliography{EnvDesignModel}

\newpage

\section{Appendix}

\subsection{Example Instance of the Environment Design Problem} \label{ExampleInstance}

A small, toy instance of the environment design problem is shown below. 

\begin{figure}[ht]
\begin{center}
\centerline{\includegraphics[width=6.5cm]{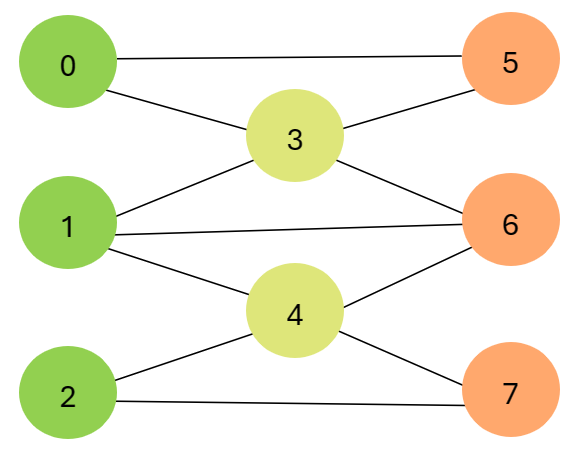}}
\caption{Example graph for the environment design problem}
\label{ExampleGraph}
\end{center}
\vskip -0.3in
\end{figure}

\begin{figure}[ht]
\begin{center}
\centerline{\includegraphics[width=6.5cm]{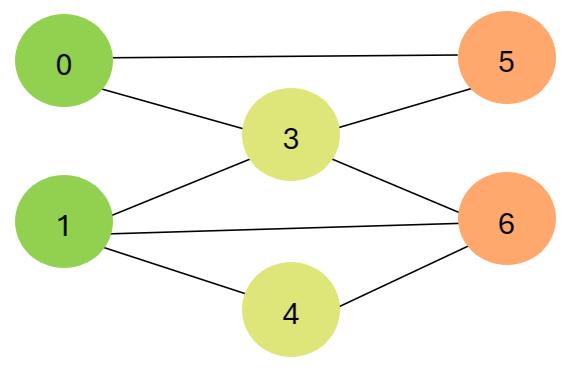}}
\caption{Example graph for the environment design problem}
\label{ExampleGraphScoped}
\end{center}
\vskip -0.3in
\end{figure}

Inputs:
\begin{itemize}
\item Undirected graph $G = (V = V_1 \cup V_2 \cup V_3, E = \{(0, 3), (0, 5), (1, 3), (1, 4), (1, 6),$ $ (2, 4), (2, 7), (3, 5), (3, 6), (4, 6), (4, 7)\})$, where $V_1 = \{0, 1, 2\}, V_2 = \{3, 4\}, V_3 = \{5, 6, 7\}$. This graph is shown in Figure \ref{ExampleGraph}.
\item Number of testing configurations $n = 3$.
\item Dimension based objective function $O(S) = $ $0.4\Sigma_{v \in V_1} \frac{(true_v - target_v)^2}{|V_1|}$ $+ 0.4\Sigma_{v \in V_2} \frac{(true_v - target_v)^2}{|V_2|}$ $+ 0.2\Sigma_{v \in V_3} \frac{(true_v - target_v)^2}{|V_3|}$, which calculates a sum of mean-squared errors between true and target distributions across all dimensions ($true_v$ and $target_v$ denote the true and target distribution values of a dimension value $v$, respectively). The original target distribution values are shown for each testing dimension in Tables \ref{Dim1Targets}, \ref{Dim2Targets}, \ref{Dim3Targets}.
\item Scope $C = (I = I_1 \cup I_2 \cup I_3, X = X_1 \cup X_2 \cup X_3)$, where $I_1 = \{0, 1\}, I_2 = \emptyset, I_3 = \emptyset, X_1 = \emptyset, X_2 = \emptyset, X_3 = \{7\}$. The scoped graph is shown in Figure \ref{ExampleGraphScoped}. The new/normalized target distribution values are shown for each testing dimension in Tables \ref{Dim1Targets}, \ref{Dim2Targets}, \ref{Dim3Targets}.
\end{itemize} 

The optimal solution to this instance of the environment design problem is $S_{opt} = [(0, 3, 5), (0, 3, 5), (1, 4, 6)]$.

$S_{opt}$ abides by all the constraints of the problem. $S_{opt}$ contains exactly $n = 3$ configurations. Every configuration in $S_{opt}$ corresponds to a clique in $G$, meaning that the dimension values in each configuration are all compatible with each other. $S_{opt}$ abides by the coverage and scope constraints. For every dimension where an include scope is provided (just $V_1$), all dimension values in that include scope are covered ($V_1 \cap I_1 = \{0, 1\}$) and all dimension values outside that include scope are not covered ($V_1 - I_1 = \{2\}$). Additionally, for every dimension where an exclude scope is provided (just $V_3$), all dimension values in that exclude scope are not covered ($V_3 \cap X_3 = \{7\}$).

$S_{opt}$ is the optimal solution since $O(S_{opt}) = 0$. $O(S_{opt}) = 0$ because the true distribution of dimension values in $S_{opt}$ matches the target distribution exactly. The true distribution values of $S_{opt}$ are shown in Tables \ref{Dim1Targets}, \ref{Dim2Targets}, \ref{Dim3Targets}.

\begin{table}[t]
\vskip 0in
\begin{center}
\begin{small}
\begin{sc}
\begin{tabular}{lccc}
\toprule
Dim. Value & Old Target & New Target & True\\
\midrule
0 & 0.6 & 0.67 & 0.67 \\
1 & 0.3 & 0.33 & 0.33 \\
2 & 0.1 & - & -\\
\bottomrule
\end{tabular}
\end{sc}
\end{small}
\end{center}
\caption{Target distribution for first testing dimension, $V_1$}
\vskip 0.1in
\label{Dim1Targets}
\end{table}

\begin{table}[t]
\vskip 0.1in
\begin{center}
\begin{small}
\begin{sc}
\begin{tabular}{lccc}
\toprule
Dim. Value & Old Target & New Target & True\\
\midrule
3 & 0.67 & 0.67 & 0.67 \\
4 & 0.33 & 0.33 & 0.33 \\
\bottomrule
\end{tabular}
\end{sc}
\end{small}
\end{center}
\caption{Target distribution for second testing dimension, $V_2$}
\vskip 0in
\label{Dim2Targets}
\end{table}

\begin{table}[t]
\vskip 0.1in
\begin{center}
\begin{small}
\begin{sc}
\begin{tabular}{lccc}
\toprule
Dim. Value & Old Target & New Target & True\\
\midrule
5 & 0.6 & 0.67 & 0.67 \\
6 & 0.3 & 0.33 & 0.33 \\
7 & 0.1 & - & -\\
\bottomrule
\end{tabular}
\end{sc}
\end{small}
\end{center}
\caption{Target distribution for third testing dimension, $V_3$}
\vskip 0in
\label{Dim3Targets}
\end{table}

\subsection{Proof of $NP$-hardness for the Environment Design Problem} \label{EDP_NPHard_Proof}

To prove that a problem is $NP$-hard, it suffices to show that an $NP$-complete problem can be reduced to that problem in polynomial time \cite{CLR4}. To prove that the environment design problem is $NP$-hard, we show a polynomial-time reduction from the $NP$-complete clique cover problem.

The inputs to the clique cover problem are a graph $G = (V, E)$ and a bound on the number of cliques $n$ \cite{Karp}. 

To convert this instance of the clique cover problem to an instance of the environment design problem, we start by creating a new graph in polynomial time, like so.

First, we create $|V|$ groups of vertices for a new graph. Each vertex group has $|V|$ vertices in it, where each vertex in that group corresponds to exactly one vertex in $V$. Each group of vertices corresponds to a dimension in the environment design problem.

For each pair of dimensions, and for each pair of vertices in those dimensions where both vertices are in different dimensions and do not correspond to the same vertex in $V$, add an edge between them if there exists an edge in $E$ between the corresponding vertices in $V$.

For each pair of dimensions, except for the pair consisting of the first and second dimensions, add edges between the vertices that correspond to the same vertex in $V$.

The resulting new graph is the input graph for the environment design problem.

In this new graph, having multiple vertices that correspond to each vertex in $V$ allows cliques of different sizes in $G$ to be represented as cliques of size $|V|$ (the number of dimensions), with one vertex in each dimension (a constraint of the environment design problem). These cliques can be used in a solution to the environment design problem and can be converted to valid cliques in $G$ by mapping each vertex of the clique to the corresponding vertex in $V$ (different vertices in the new graph corresponding to the same vertex in $V$ are consolidated by being mapped to that one vertex).

Having one pair of dimensions without the edges between the vertices corresponding to the same vertex in $G$ ensures that no cliques where all the vertices map to the same vertex in $V$ are used in a solution to the environment design problem. Such cliques would correspond to single vertices in $G$ and are not valid cliques that can be used in a solution to the clique cover problem.

Having $|V|$ dimensions ensures that cliques of any size in $G$ (no clique in $G$ can have more than $|V|$ vertices) can be used in the solution to the environment design problem. 

Another input to the environment design problem is the number of testing configurations, which is the same as the number of cliques in the clique cover problem, $n$.

For this proof, the optimality of a solution to the environment design problem is not important, as long as the constraints are satisfied. So, for this proof, a trivial objective function (one that always returns 0) can be an input to the environment design problem.

The exclude scope is empty. The include scope consists of one vertex from each dimension, each vertex corresponding to a different vertex in $V$. This ensures that all vertices in $V$ are covered in any solution to the environment design problem.

A solution to the environment design problem, a list of $n$ cliques (each clique containing $|V|$ vertices), can be converted into a solution for the clique cover problem in polynomial time. For each clique in the solution to the environment design problem, map each vertex in the clique to the corresponding vertex in $V$. The resulting set of mapped vertices represents a valid clique in $G$. The list of $n$ cliques produced by this process is a valid clique cover of $G$, and therefore is a valid solution to the environment design problem.

Hence, the $NP$-complete clique cover problem is reducible in polynomial time to the environment design problem, proving that the environment design problem is $NP$-hard. 

\end{document}